# The Problem of Colliding Networks and its Relation to Cancer

Alexei A. Koulakov and Yuri Lazebnik

*Cold Spring Harbor Laboratory, Cold Spring Harbor, NY 11724*

Complex systems, ranging from living cells to human societies, can be represented as attractor networks, whose basic property is to exist in one of allowed states, or attractors. We noted that merging two systems that are in distinct attractors creates uncertainty, as the hybrid system cannot assume two attractors at once. As a prototype of this problem, we explore cell fusion, whose ability to combine distinct cells into hybrids was proposed to cause cancer. By simulating cell types as attractors, we find that hybrids are prone to assume spurious attractors, which are emergent and sporadic states of networks, and propose that cell fusion can make a cell cancerous by placing it into normally inaccessible spurious states. We define basic features of hybrid networks and suggest that the problem of colliding networks has general significance in processes represented by attractor networks, including biological, social, and political phenomena.

A cell can be viewed as a network that interrelates tens of thousands of genes and an even larger inventory of their products. The presence of about 400 cell types in the human body(1) implies that this network can assume at least as many distinct, stable, and self-maintaining states. A basic and largely unexplored question is what happens if these distinct states collide during cell fusion, a fundamental yet poorly understood biological process that merges two or more cells into one(2-5). This question is of particular interest, because cell fusion is a central force in evolution(6, 7), has been used in therapy(8, 9), and was proposed to cause cancer(10-13).

Cell fusion in the body is normally restricted to a few physiological processes, is tightly controlled, and occurs only between cells of particular types, such as between the sperm and the egg or between muscle precursors(2). However, cells that are not supposed to fuse do so occasionally, particularly during common viral infections(10). The fate of the resulting hybrids is largely unknown, but recapitulating their creation in the laboratory suggests that hybrids and cancer cells share a set of properties. Both cancer cells and some hybrids give rise to tumors and can assume diverse abnormal cell types(3, 14). The diversity of cancerous cells is such that it is common even for experienced pathologists to disagree on identifying a particular specimen(15). The origin of this heterogeneity is not entirely clear. Products of cell fusion are also diverse, as evidenced by genome-wide gene expression studies(3, 14), with the mechanism of this diversity remaining to be completely understood.

The cause of the diversity among hybrids might be an underappreciated intrinsic feature of cell fusion, a conflict resulting from fusing distinct cells. The main cause of this conflict may be contradiction between the parental self-maintaining network states that underlie cell types. For example, fusing an epithelial cell and a macrophage, which is a combination implicated in carcinogenesis(10), may result in a hybrid that is an average of its parents, is more similar to one of them, or is unlike either of them(4, 5). The rules determining this choice are unclear(3). It is reasonable to suggest that the conflict between the states would continue until they reach a consensus, one of them wins, or the cell dies in the process due to irreconcilable differences. We will call the problem of reaching a consensus between distinct self-maintaining states the problem of colliding networks.

To understand possible outcomes of cell fusion, we simulated this process computationally. We assumed that a cell type is a stable state of a cell-autonomous gene interaction network. Therefore, it can be viewed as a network attractor(16-18), which is a state that is stable with respect to small perturbations and noise (Fig. 1 A). Cell types as attractors are commonly visualized using the concept of epigenetic space(16, 18). Each point in this space represents a particular state of the cell. This state is defined, for example, by the pattern of gene activities and described by an energy-like function that can be visualized by the altitude of a cell residing on the epigenetic Waddington landscape. The landscape has multiple basins, the bottom points of which are attractors. Each attractor represents a state with a locally minimal energy in the attractor's basin. As a result, a cell deflected from this state will spontaneously return to it unless it was deflected beyond the edges of the attractor's basin.

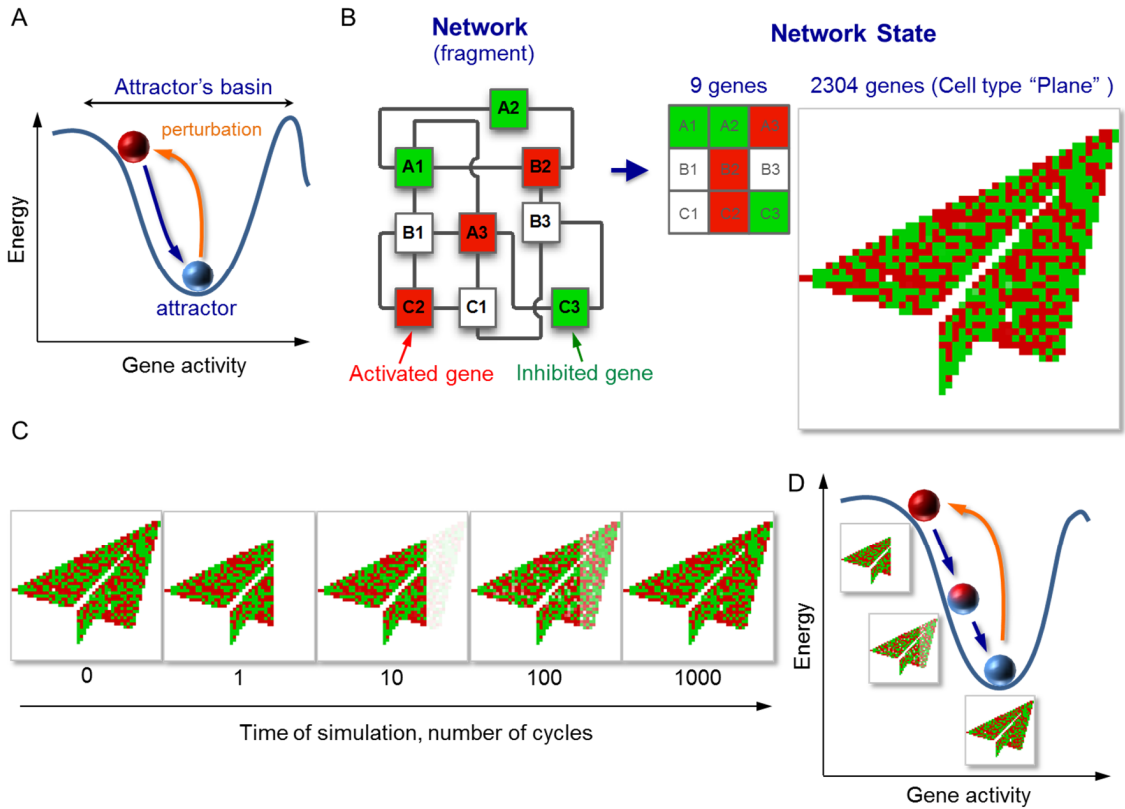

**Figure 1. Cell types represented as network attractors.** (A) An attractor is a state of a network in which the energy of this network is locally minimal within the region known as the basin of the attractor. The current state of a cell is represented as a ball at the bottom of the attractor. The ball displaced within the attractor's basin returns to the attractor. (B) Cell types modeled as network attractors. A fragment of the network used in this study illustrates that it is composed of interconnected nodes (genes A1 to B3). The state of this network is represented by an array in which activated genes are indicated in red, inhibited genes in green, and genes with basal activity in white. The entire network contains 2304 ($48^2$) genes and thus its state is visualized by a 48 x 48 array. The network is preprogrammed to contain a set of attractor states (embedded cell types) that correspond to a set of recognizable pictorial symbols, such as Plane. (C, D) To verify that the Plane is an attractor state, the nose of the Plane is cut by resetting a group of the genes to their basal activity. The network spontaneously restores the "nose" by returning in its initial state, thus confirming that this state is an attractor.

To represent cell types as attractors, we adopt arguably the most studied model of attractor networks, a mathematical formalism introduced by Hopfield to simulate associative memory(19, 20). Our implementation includes several descriptors (Fig. 1B). First, we define the network nodes, which we call "genes." Each gene can be activated, inhibited, or retain its basal activity. The pattern of activities of all of the genes defines the network state. In the mathematical sense, this pattern is represented as a vector with the length equal to the number of genes. If a gene in the network is activated or suppressed, the corresponding entry in the vector becomes positive or negative respectively. This vector can be visualized as an array (Fig. 1B) in which each unit represents a gene. In this array, activated genes are denoted in red, inhibited genes in green, and genes whose activity is basal in white. Because our network uses $48^2$ (2304) genes, the state of this network is visualized by a 48 x 48 array. Assuming that each cell type in the body has a specific pattern of gene expression, we represent cell types in the model by a set of gene activity arrays that are easily recognizable as pictorial symbols (Fig. 1B).

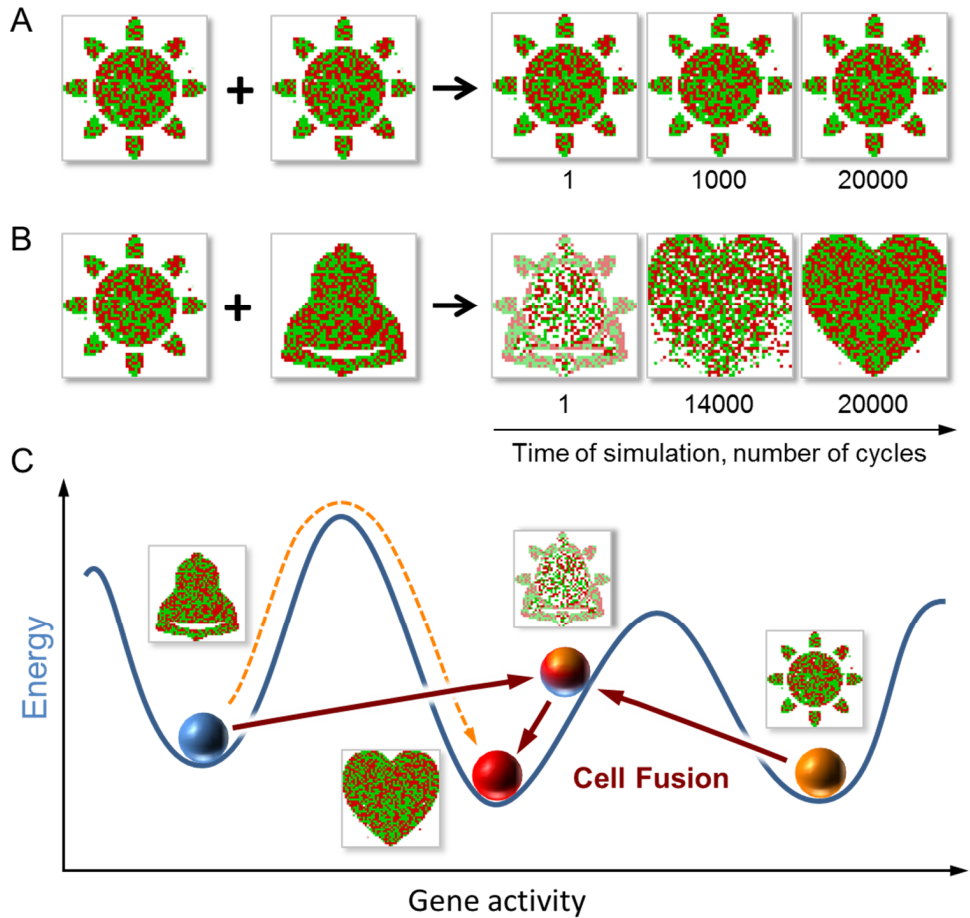

**Figure 2.** Simulated cell fusion can change cell types. (A) Fusion of two cells of the same type leads to a hybrid of the same type. (B) Fusion of two cell types (Sun and Bell) results in a hybrid that belongs to another cell type (Heart). (C) Interpretation of (B). Cell fusion can transport a cell from one attractor (Bell) to the basin of another (Heart) by creating an intermediate product (Bell+Sun) with another cell (Sun). This mechanism bypasses the need to overcome the hills of the epigenetic space.

Finally, we define interactions between the genes. These interactions can be positive or negative, reflecting the fact that genes can activate or suppress each other. Following the Hopfield model(19, 20), we assume that the strength of interaction between any two genes is proportional to the incidence with which their activities correlate in all gene activity patterns that we designated as cell types [Methods, equation (2)]. Therefore, the more often any two genes are co-active, the stronger they activate each other in the model. Similarly, the more often the activities of any two genes anti-correlate, the stronger these genes inhibit each other. All gene interactions are assumed to be symmetric, meaning that if gene A can activate gene B, then the reverse is also true for simplicity. A key property of Hopfield networks is that if the interactions within every pair of genes are set as described, the patterns used to calculate these values become network attractors. Therefore, the pictorial symbols that we designate as cell types also become network attractors, which we call "embedded cell types."

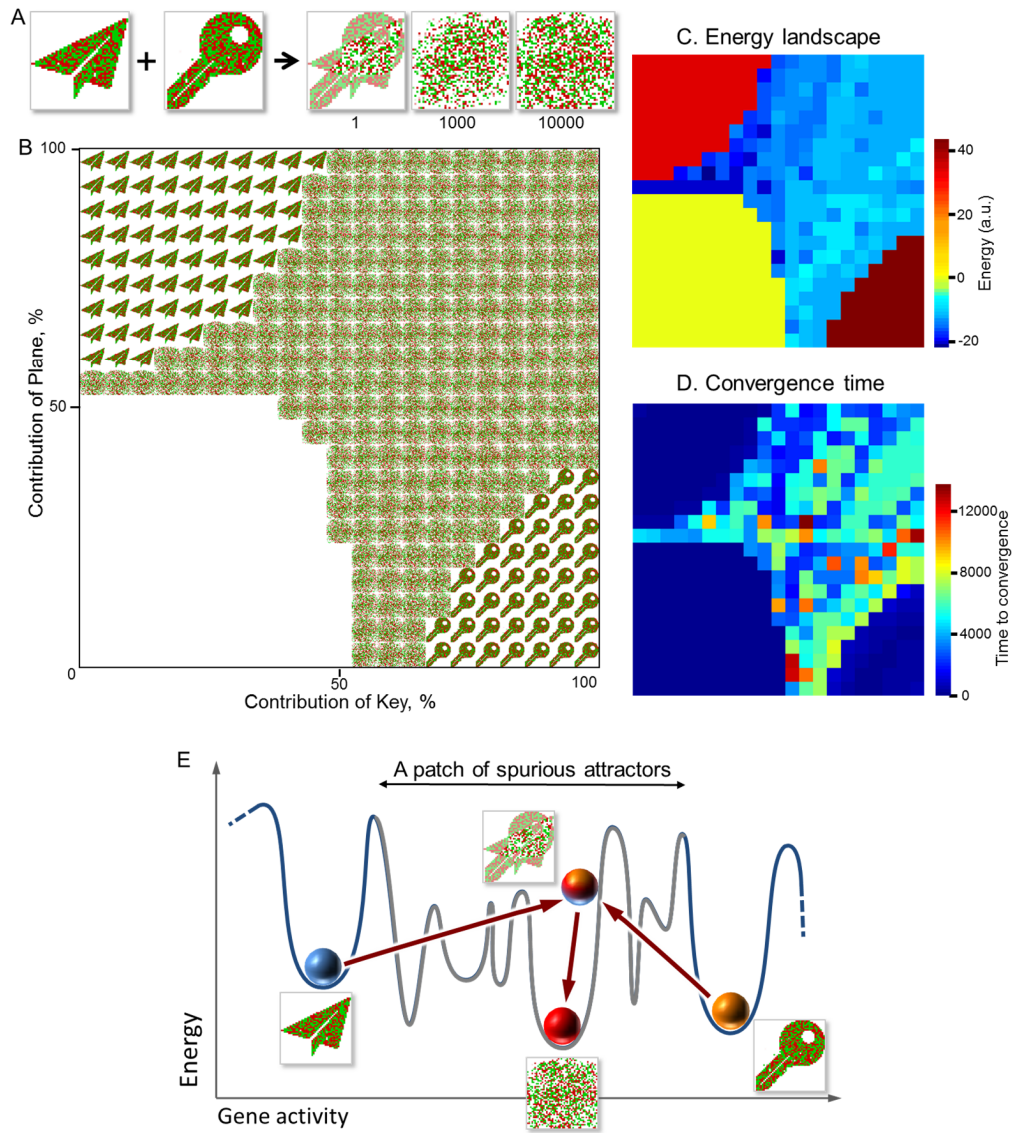

**Figure 3.** Simulated fusion of two embedded cell types places the resulting cell into one of numerous spurious states. (A) The evolution of the hybrid between Plane and Key. (B) Fusing two cell types results in a set of diverse hybrids. The panel shows the final settled gene activity patterns of hybrids obtained by fusing Plane and Key with varied contribution of each of these cell types at 5% increments. The white quadrant (bottom left) represents the states in which all genes acquire their basal activity, an equivalent of cell death. (C) Spurious states differ in their energy. The panel shows the energy map for hybrids that are displayed in (B). 3-D representation of this map is shown in Supplemental Fig. 1. Spurious attractors (the middle of the image) are heterogeneous as manifested by the diverse colors on the map and their depth. (D) Settling into spurious states takes longer than into the embedded states. The image shows the average time to reach the final states displayed in (B). A 3-D representation is in Supplementary Figure 2. (E) Cell fusion can place cells into spurious attractors. Fusion of Plane to Key creates a product that is in the basin of a spurious attractor. The abundance of spurious attractors (Fig. 3B-D) suggests that they can exist in patches that cover the space between embedded (normal) attractors.

To verify that embedded cell types are attractors, we perturb one of them ("Plane") by resetting some of its genes to their basal state (note the missing "nose" of the Plane in Fig 1C, "T=1"). As expected, the network automatically restores the silenced region (see Fig. 1C), confirming that Plane is an attractor.

Our model included 90 embedded cell types, which is comparable to the number of cell types found in humans.

To simulate cell fusion, we assume that the gene activity pattern (i.e. the network state) of the hybrid is initially an average of the parental states. In other words, the activity of a gene in the newly formed hybrid is the average of the activities of this gene in the parents. We also assume that fusion does not change how the genes interact with each other. Once the hybrid network is formed, the gene activity pattern evolves according to the rules of gene interactions. To test our simulation, we first fuse two identical cell types. Fusing two Sun cells results in a hybrid of the same type (Fig. 2A). This implies that in the absence of differences between the parental states, the hybrid remains in the same attractor as its parents.

Fusing different cell types produces two types of hybrids. In one case, the hybrids assume an embedded cell type. For example, fusing Sun and Bell results in Heart (Fig. 2B). We interpret this result by assuming that the immediate product of combining Sun and Bell landed in the basin of the attractor Heart, because this attractor is located in the epigenetic space somewhere between the parental attractors (Fig. 2C). This interpretation implies that cell fusion can almost instantly move a cell from attractor A to the basin of attractor B by creating an intermediate product with another cell. Because this process bypasses the need to overcome the hills that can separate A and B in the epigenetic landscape, one can argue that cell fusion allows a cell to hop between two attractors with the help of another cell.

Another class of hybrid networks assumes stable self-maintaining states that are unlike any of the embedded cell types. For example, fusing Plane to Key (Fig. 3A) results in such a *de novo* cell type. Such states are known in network theory as spurious attractors(20, 21). They are emergent, unavoidable, and usually undesirable features of complex networks. Thus, we assume that the hybrid of Plane and Key happened to be in the basin of a spurious attractor.

To test how abundant the spurious attractors are, we systematically change the conditions of fusion by varying the relative contributions of the parents, Plane and Key (Fig. 3B). If the contribution of one parent was relatively large, the hybrid assumed the cell type of this parent (Fig. 3B, top left and bottom right quadrants of the fusion table). This behavior can be envisioned as a tug-of-war between two attractors, in which weakening one competitor gives victory to the other. If the contribution of both parents was below a certain threshold, then all genes in the hybrids assume their basal activity (Fig. 3B, bottom left quadrant of the fusion table, white arrays). This event could be interpreted as an equivalent of cell death, implying that cell death is one of the attractors. The intermediate range of parental contributions produces numerous spurious states (Fig 3B, center). Notably, even an incremental change in parental contribution places the hybrid into a spurious attractor, implying that the basins of embedded and spurious attractors are close to each other.

To determine whether spurious states are diverse, we compare their energy defined as the Lyapunov function of the network (Fig. 3C and Supplementary Methods). Spurious states differ by their energy, indicating that they are indeed diverse. The multitude and diversity of spurious attractors accessible to the hybrids of two cell types suggest that the number of spurious states in the network exceeds the number of embedded states.

Reaching spurious states in our system takes much longer than returning to the parental attractor (Fig. 3D). This means that a newly formed hybrid system can undergo a long set of transformations before settling into its final spurious state. Our findings suggest that cell fusion can initiate a long chain of spontaneously occurring transformations. We also find that the convergence to the spurious state is not gradual and is punctuated by sudden changes in the network state separated by the long periods of relative stability (Figure S3).

Our model makes several predictions for the behavior of merged attractor networks. First, the products of network collision can assume not only the predetermined embedded states but also abnormal states.

Second, the number of states that a hybrid network can assume exceeds substantially the number of predetermined attractor states. Third, the final state of the hybrid system can be unpredictable and depend on incremental changes in fusion conditions. Fourth, the hybrid system can spontaneously evolve following fusion over extended periods of time that exceed settling to the 'normal' embedded state. Finally, this evolution is not gradual as it can be punctuated by the sudden changes in the network state.

These predictions are consistent with what is known about cell fusion and with the model that cell fusion can cause cancer. Cell fusion can produce both normal and abnormal cells(2). The products of fusion are not mere averages of the parental cell types(4, 5) but rather that each hybrid is unique(3, 14, 22-29), as our model predicts (Figures 2B and 3A). Similarly, multiplicity of spurious states in our system is reminiscent of the diversity that hybrid cells can assume(3, 14, 22-29). Our model also suggests that the patterns of gene expression in hybrids change over time before settling into their final state (Figure 3), as experimentally observed(3). If the ability to place cells into abnormal states is indeed a basic property of cell fusion, then using cell fusion as a therapeutic tool(8), particularly in stem cell therapy(9), should be considered with an abundance of caution.

In our study we considered a simplified version of intracellular network. Our model does not include compartmentalization of gene products within a cell, asymmetry in network interactions, interactions that involve more than two products, particular properties of eukaryotic genes and proteins, the dynamic nature of cellular attractors, cell-cell signaling, etc. We also assumed that cell fusion does not affect gene-gene interactions. An open question is whether incorporating these parameters will lead to a substantial change in our conclusions. We surmise that these phenomena will lead to a more complex set of behaviors that are more difficult to reconcile after cell fusion.

Previous models suggest that cells become cancerous by assuming abnormal attractors(30, 31). The origins of these attractors and how the cells get there is unclear. We propose that abnormal attractors are the spurious states inherent to attractor networks(20, 21) (Fig. 3E). Cell fusion can help a cell reach attractors and thus assume corresponding cell types that are otherwise inaccessible. This property is beneficial if the final attractor corresponds to a normal cell type, such as osteoclast or a myofiber. However, given the abundance of spurious states in complex attractor networks(20, 21), accidental fusion may lead to the emergence of hybrids with abnormal cell types, including those that share properties with cancer cells. Our observations suggest that spontaneous behavior of hybrid networks is consistent with two basic properties of carcinogenesis. The protracted behavior of networks converging to spurious attractors is similar to the dynamics of carcinogenesis(32), while the abundance and diversity of the spurious attractors is reminiscent of the diversity of cancerous cells(33).

The behavior of hybrid networks in our model is also reminiscent of mergers of large businesses, organizations that have been considered complex attractor networks(34). Business mergers can produce viable companies, which are predetermined intended final states of the hybrid networks, but can also result in companies that fail. Despite ongoing changes introduced by the management and leadership in response to emerging problems, the number of failed business mergers exceeds the number of those that produce the desired result(35). We suggest that the similarities between the consequences of merging living cells and businesses are not superficial, but may reflect basic properties of hybrid networks.

Because our simulation considers network elements and their interactions as abstractions, the implication of our study is that the propensity to assume abnormal states might be a fundamental emergent property of colliding networks, rather than a result of how a particular network is implemented. As a result, combining complex attractor networks, whether they are cells, species, personalities, businesses, or countries is inherently fraught with the possibility of unintended and lasting outcomes.

**METHODS**

To model the intracellular regulatory network, we use time-dependent network equations similar to the continuous Hopfield model(20):

$$\tau dx_i / dt = u_i - x_i, \qquad (1)$$

where $\tau$, $x_i$, and $u_i$ are the time-constant, the concentration of products of gene number $i$ in the cell, and the rate of its production, respectively. Variable $x_i$ is the activity vector that is shown in Figures 1 through 3 show a colored square array. The rate of gene production is related to the concentration of other products by the network interaction matrix $u_i = F\left(\sum_j W_{ij} x_j\right)$. Here, $F(a)$ is a sparse (36, 37) saturating non-linear function that relates the rate of gene production to the concentrations of other gene products. We use $F(a) = \tanh[(a-\theta)/\lambda]$ for $a \geq \theta$, $F(a) = \tanh[(a+\theta)/\lambda]$ for $a \leq -\theta$, and zero otherwise ($\lambda = 0.1$, $\theta = 0.5$). This form of non-linearity makes our model behave similarly to the models of sparse overcomplete representations (36, 37) and compressive sensing approaches to signal processing (38). The network weight matrix contains information about the patterns of gene activities in embedded cell types:

$$W_{ij} = \sum_c \xi_i^c \xi_j^c / N^c. \qquad (2)$$

Here, $\xi_i^c$ is the gene activity pattern corresponding to cell type number $c$, $N^c$ is the number of non-zero pixels in this pattern, and the sum is assumed over embedded 90 cell types. To implement cell fusion, we form the combination of gene activities corresponding to attractors from two parent cells $x_i^{hybrid} = c_1 x_i^{parent_1} + c_2 x_i^{parent_2}$. The coefficients $c_1$ and $c_2$ determine the contributions of the parents and vary from 0 to 1. This pattern is used as an initial condition for Equation (1). The pattern Heart was adjusted to result from fusion of Sun and Bell, as detailed in Supplementary Methods. In the Supplementary methods we show that the dynamics of the network can be viewed as a gradient descent on the landscape defined by the Lyapunov function. This function plays the role of energy in our model (36, 37). To evaluate the convergence time to the final configuration, we calculated change in the gene activity vector as a function of time $\Delta x(t) = \sum_i |x_i(t+1) - x_i(t)|$. The time to convergence after fusion is calculated as $\bar{t} = \sum_t t \cdot \Delta x(t) / \sum_t \Delta x(t)$. This variable is displayed in Figure 3D. Our methods are described in more detail in the Supplementary Materials.

# REFERENCES

1. Vickaryous MK & Hall BK (2006) Human cell type diversity, evolution, development, and classification with special reference to cells derived from the neural crest. *Biol Rev Camb Philos Soc* 81(3):425-455.
2. Oren-Suissa M & Podbilewicz B (2010) Evolution of programmed cell fusion: common mechanisms and distinct functions. *Dev Dyn* 239(5):1515-1528.
3. Palermo A*, et al.* (2009) Nuclear reprogramming in heterokaryons is rapid, extensive, and bidirectional. *Faseb J* 23(5):1431-1440.
4. Massa S, Junker S, & Matthias P (2000) Molecular mechanisms of extinction: old findings and new ideas. *Int J Biochem Cell Biol* 32(1):23-40.
5. Bulla GA, Luong Q, Shrestha S, Reeb S, & Hickman S (2010) Genome-wide analysis of hepatic gene silencing in mammalian cell hybrids. *Genomics* 96(6):323-332.
6. Leitch AR & Leitch IJ (2008) Genomic plasticity and the diversity of polyploid plants. *Science* 320(5875):481-483.
7. Mallet J (2007) Hybrid speciation. *Nature* 446(7133):279-283.
8. Gong J, Koido S, & Calderwood SK (2008) Cell fusion: from hybridoma to dendritic cell-based vaccine. *Expert Rev Vaccines* 7(7):1055-1068.
9. Yamanaka S & Blau HM (2010) Nuclear reprogramming to a pluripotent state by three approaches. *Nature* 465(7299):704-712.
10. Duelli D & Lazebnik Y (2007) Cell-to-cell fusion as a link between viruses and cancer. *Nat Rev Cancer* 7(12):968-976.
11. Lu X & Kang Y (2009) Cell fusion as a hidden force in tumor progression. *Cancer Res* 69(22):8536-8539.
12. Bjerkvig R, Tysnes BB, Aboody KS, Najbauer J, & Terzis AJ (2005) Opinion: the origin of the cancer stem cell: current controversies and new insights. *Nat Rev Cancer* 5(11):899-904.
13. Matsuura K*, et al.* (2011) Identification of a link between Wnt/beta-catenin signalling and the cell fusion pathway. *Nat Commun* 2:548.
14. Ambrosi DJ*, et al.* (2007) Genome-wide reprogramming in hybrids of somatic cells and embryonic stem cells. *Stem Cells* 25(5):1104-1113.
15. Allison KH*, et al.* (2008) Diagnosing endometrial hyperplasia: why is it so difficult to agree? *Am J Surg Pathol* 32(5):691-698.
16. Macarthur BD, Ma'ayan A, & Lemischka IR (2009) Systems biology of stem cell fate and cellular reprogramming. *Nat Rev Mol Cell Biol* 10(10):672-681.
17. Kauffman SA (1993) *The origins of order : self-organization and selection in evolution* (Oxford University Press, New York ; Oxford) pp xviii, 709 p.
18. Bar-Yam Y, Harmon D, & de Bivort B (2009) Systems biology. Attractors and democratic dynamics. *Science* 323(5917):1016-1017.
19. Hopfield JJ (1982) Neural Networks and Physical Systems with Emergent Collective Computational Abilities. *P Natl Acad Sci-Biol* 79(8):2554-2558.
20. Hertz J, Krogh A, & Palmer RG (1991) *Introduction to the theory of neural computation* (Westview, a Member of the Perseus Books Group, Cambridge, Mass.) pp xxii, 327 p.
21. Amit DJ, Gutfreund H, & Sompolinsky H (1985) Storing Infinite Numbers of Patterns in a Spin-Glass Model of Neural Networks. *Physical Review Letters* 55(14):1530-1533.
22. Powell AE*, et al.* (2011) Fusion between Intestinal epithelial cells and macrophages in a cancer context results in nuclear reprogramming. *Cancer Res* 71(4):1497-1505.
23. Cowan CA, Atienza J, Melton DA, & Eggan K (2005) Nuclear reprogramming of somatic cells after fusion with human embryonic stem cells. *Science* 309(5739):1369-1373.

24. Mukhopadhyay KD, *et al.* (2011) Isolation and characterization of a metastatic hybrid cell line generated by ER negative and ER positive breast cancer cells in mouse bone marrow. *PLoS One* 6(6):e20473.
25. Dittmar T, *et al.* (2011) Characterization of hybrid cells derived from spontaneous fusion events between breast epithelial cells exhibiting stem-like characteristics and breast cancer cells. *Clin Exp Metastasis* 28(1):75-90.
26. Wang Y, *et al.* (2011) Fusion of human umbilical cord mesenchymal stem cells with esophageal cells. *Int J Oncol*.
27. Lu X & Kang Y (2009) Efficient acquisition of dual metastasis organotropism to bone and lung through stable spontaneous fusion between MDA-MB-231 variants. *Proc Natl Acad Sci U S A* 106(23):9385-9390.
28. Klee WA & Nirenberg M (1974) A neuroblastoma times glioma hybrid cell line with morphine receptors. *Proc Natl Acad Sci U S A* 71(9):3474-3477.
29. Ringertz NR & Savage RE (1976) *Cell hybrids* (Academic Press, New York) pp xiv, 366 p.
30. Huang S, Ernberg I, & Kauffman S (2009) Cancer attractors: a systems view of tumors from a gene network dynamics and developmental perspective. *Semin Cell Dev Biol* 20(7):869-876.
31. Kaneko K (2011) Characterization of stem cells and cancer cells on the basis of gene expression profile stability, plasticity, and robustness: dynamical systems theory of gene expressions under cell-cell interaction explains mutational robustness of differentiated cells and suggests how cancer cells emerge. *Bioessays* 33(6):403-413.
32. Feinberg AP, Ohlsson R, & Henikoff S (2006) The epigenetic progenitor origin of human cancer. *Nat Rev Genet* 7(1):21-33.
33. Heng HH, *et al.* (2011) Evolutionary mechanisms and diversity in cancer. *Adv Cancer Res* 112:217-253.
34. Wilkinson I & Young L (2002) On cooperating: firms, relations and networks. *Journal of Business Research* 55(2):123-132.
35. Bruner RF (2005) *Deals from hell : M & A lessons that rise above the ashes* (Wiley, Hoboken, N.J.) pp xii, 420 p.
36. Rozell CJ, Johnson DH, Baraniuk RG, & Olshausen BA (2008) Sparse coding via thresholding and local competition in neural circuits. *Neural Comput* 20(10):2526-2563.
37. Koulakov AA & Rinberg D (2011) Sparse incomplete representations: a potential role of olfactory granule cells. *Neuron* 72(1):124-136.
38. Baraniuk RG (2007) Compressive sensing. *Ieee Signal Proc Mag* 24(4):118.

## SUPPLEMENTARY METHODS

To model the intracellular regulatory network, we use time-dependent network equations similar to the continuous Hopfield model (*1*):

$$\tau dx_i / dt = u_i - x_i, \quad (1)$$

where $\tau$, $x_i$, and $u_i$ are the time-constant, the concentration of products of gene number $i$ in the cell, and the rate of its production, respectively. Variable $x_i$ is the activity vector that is shown in Figures 1 through 3 are a colored square array. The rate of gene production is related to the concentration of other products by the network interaction matrix $u_i = F\left(\sum_j W_{ij} x_j\right)$. Here, $F(a)$ is a sparse saturating non-linear function that relates the rate of gene production to the concentrations of other gene products. This function has an interval of inputs within which it is zero. This property makes it possible that some genes remain inactive, i.e. gene activities are sparse (*2, 3*). We use the non-linear function frequently employed in neural networks forming sparse representations (*2, 3*). We assumed that $F(a) = \tanh[(a-\theta)/\lambda]$ for $a \geq \theta$, $F(a) = \tanh[(a+\theta)/\lambda]$ for $a \leq -\theta$, and zero otherwise ($\lambda = 0.1$, $\theta = 0.5$). By introducing the activity of the promoter of gene $i$, $a_i \equiv \sum_j W_{ij} x_j$, we can rewrite Equation (1) as

$$\tau da_i / dt = \sum_j W_{ij} u_i - a_i, \quad (2)$$

With the additional constraint of $u_i = F(a_i)$. Equation (2) is a continuous Hopfield model (*1*). The role of energy in this equation is played by the Lyapunov function

$$L(\vec{u}) = -\frac{1}{2} \sum_j u_i W_{ij} u_j + \sum_i C(u_i). \quad (3)$$

This means that if the weight matrix $W_{ij}$ is symmetric, the system of differential equations (2) can be represented as a gradient descent with the energy function given by the Lyapunov function $\tau da_i / dt = -\partial L / \partial u_i$. Consequently, the Lyapunov function is non-increasing in our simulations, i.e. $dL/dt \leq 0$. The cost function in Equation (3) is defined by the standard expression $C(u) = \int_0^u F^{-1}(u') du'$ [c.f. Ref. (*3*)] and can be evaluated as

$$C(u) = \theta |u| + \lambda[\operatorname{atanh}(u) + \ln(1-u^2)/2] \quad (4)$$

Because of the absolute value present in the first term of the cost-function, the model behaves similarly to the models with L1-norm cost, i.e. yields sparse activity vectors with a large number of inactive genes. The Lyapunov function $L$ for the pattern of gene activities after convergence is displayed in Figure 3C.

The network weight matrix contains information about the patterns of gene activities in embedded cell types:

$$W_{ij} = \sum_c \xi_i^c \xi_j^c / N^c. \qquad (5)$$

Here, $\xi_i^c$ is the gene activity pattern corresponding to cell type number $c$, $N^c$ is the number of non-zero pixels in this pattern, and the sum is assumed over embedded 90 cell types. This expression is similar to the standard form used in the Hopfield model (*4*) with the normalization dependent on the number of active genes. The patterns were generated from a library of black and white 48x48 icons $b_i^c = 0,1$ modulated by random white-noise patterns $w_i^c = \pm 1$ through $\xi_i^c = b_i^c w_i^c$. We use a set of recognizable icons $b_i^c = 0,1$ for illustration purposes. All patterns were downloaded from an internet database of black and white computer icons.

Physiological cell fusion is a highly controlled event with predictable outcome. The identities of parental and hybrid cell types are predetermined. For example, only certain cell types can fuse to create muscle fibers. The network mechanisms that make fusion predictable are likely to emerge in the course of evolution concurrently with cell types themselves. Although we observed that some embedded cell types emerged as the result of fusion of two other embedded cell types spontaneously, we found that it is also possible to program the outcome of cell fusion by adjusting the gene activity patterns corresponding to the desired hybrid. Thus, pattern "Heart" in Figure 2B was programmed to facilitate its production from the sum of "Sun" and "Bell." This programming was performed before the fusion of these two cell types was modeled computationally. To this end, we calculated the sum of the latter two patterns and assigned the white noise variable $w_i^{heart}$ to match the sign of $w_i^{sun} + w_i^{bell}$ for 40% of pixels chosen randomly within the overlap of these two patterns.

To implement cell fusion, we form the combination of gene activities corresponding to attractors from two parent cells $x_i^{hybrid} = c_1 x_i^{parent_1} + c_2 x_i^{parent_2}$. The coefficients $c_1$ and $c_2$ determine the contributions of the parents and vary from 0 to 1. These two coefficients form the parameters for Figures 3B D. Although for conserved volume and interpreting variables $x_i$ as concentrations we expect that $c_1 + c_2 = 1$, corresponding to the diagonal in the figures, we explored a 2D range of parameters. We assume therefore that volume may not be conserved in fusion and that gene products may be localized within cells in regions with non-conserved volume. The new variable $x_i^{hybrid}$ is used as the initial condition for Equation (1). After fusion, we run a simulation for 50,000 steps. The time constant in Equation (1) is $\tau = 100$. We use simple finite differences with time step $\Delta t = 1$ to model dynamics. We verify that the pattern of gene activities reaches convergence by ensuring that the activity vector did not change substantially at the end of simulation.

To evaluate the convergence time to the final configuration, we calculate change in the gene activity vector as a function of time $\Delta x(t) = \sum_i |x_i(t+1) - x_i(t)|$. The time to convergence after fusion is calculated as $\bar{t} = \sum_t t \cdot \Delta x(t) / \sum_t \Delta x(t)$. This variable is displayed in Figure 3D.

**REFERENCES**

1. J. Hertz, A. Krogh, R. G. Palmer, *Introduction to the theory of neural computation*. Santa Fe Institute studies in the sciences of complexity. Lecture notes v. 1 (Westview, a Member of the Perseus Books Group, Cambridge, Mass., 1991), pp. xxii, 327 p.
2. C. J. Rozell, D. H. Johnson, R. G. Baraniuk, B. A. Olshausen, Sparse coding via thresholding and local competition in neural circuits. *Neural Comput* **20**, 2526 (Oct, 2008).
3. A. A. Koulakov, D. Rinberg, Sparse incomplete representations: a potential role of olfactory granule cells. *Neuron* **72**, 124 (Oct 6, 2011).
4. J. J. Hopfield, Neural Networks and Physical Systems with Emergent Collective Computational Abilities. *P Natl Acad Sci-Biol* **79**, 2554 (1982).

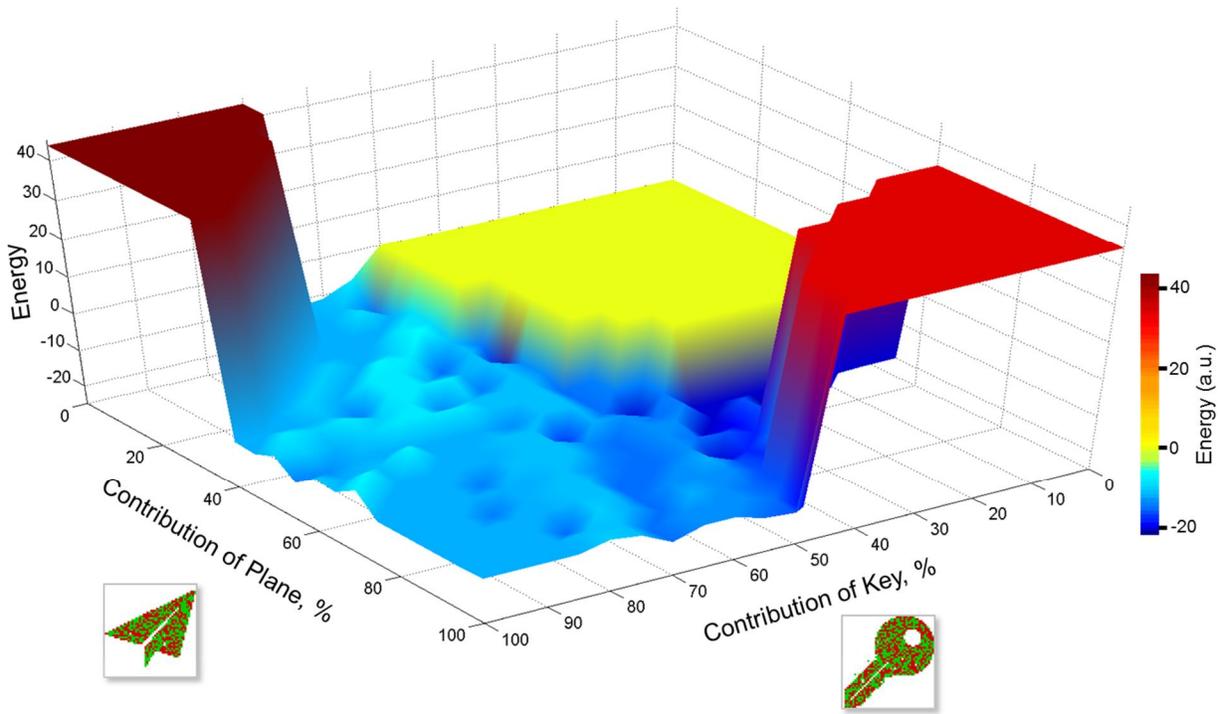

**Supplementary Figure 1.** Energy landscape of hybrids resulting from fusion of Plane and Key. This is a 3D representation of Figure 3C. The energy is given by the Lyapunov function defined in equation (3).

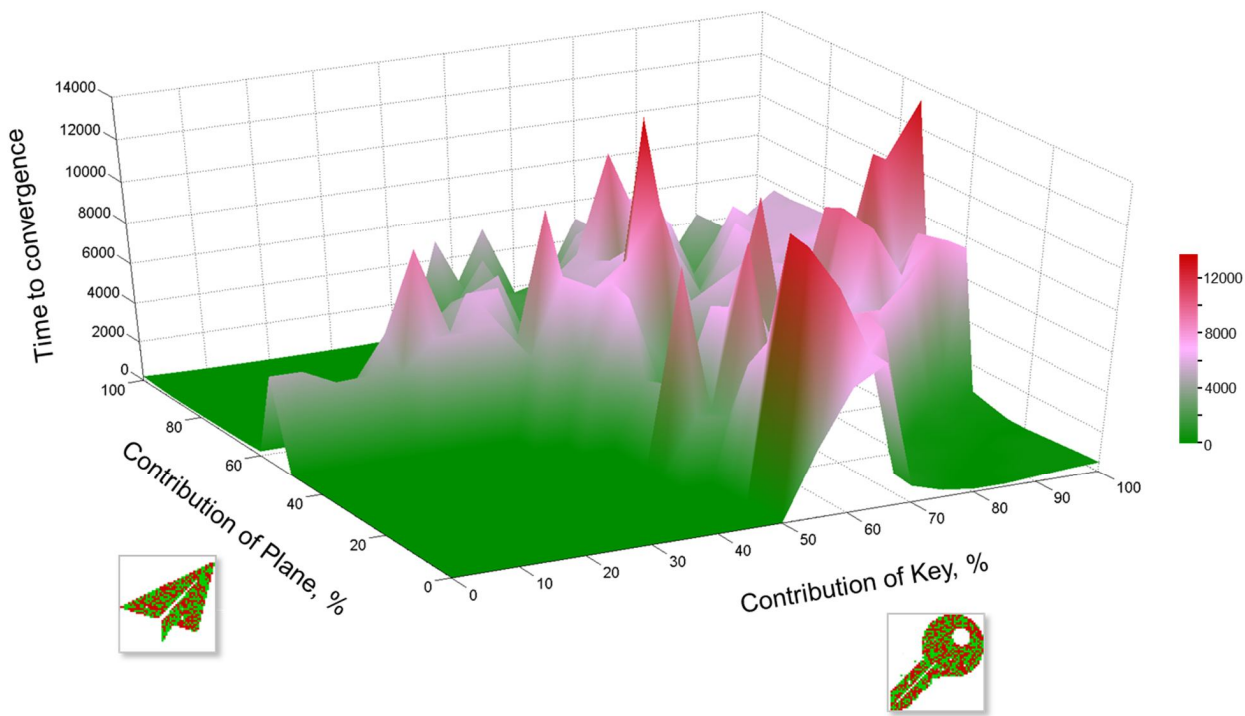

**Supplementary Figure 2.** Time to convergence for hybrids resulting from fusion of Plane and Key. This is a 3D representation of Figure 3D. Time to convergence is defined in the last paragraph of Supplementary Methods.

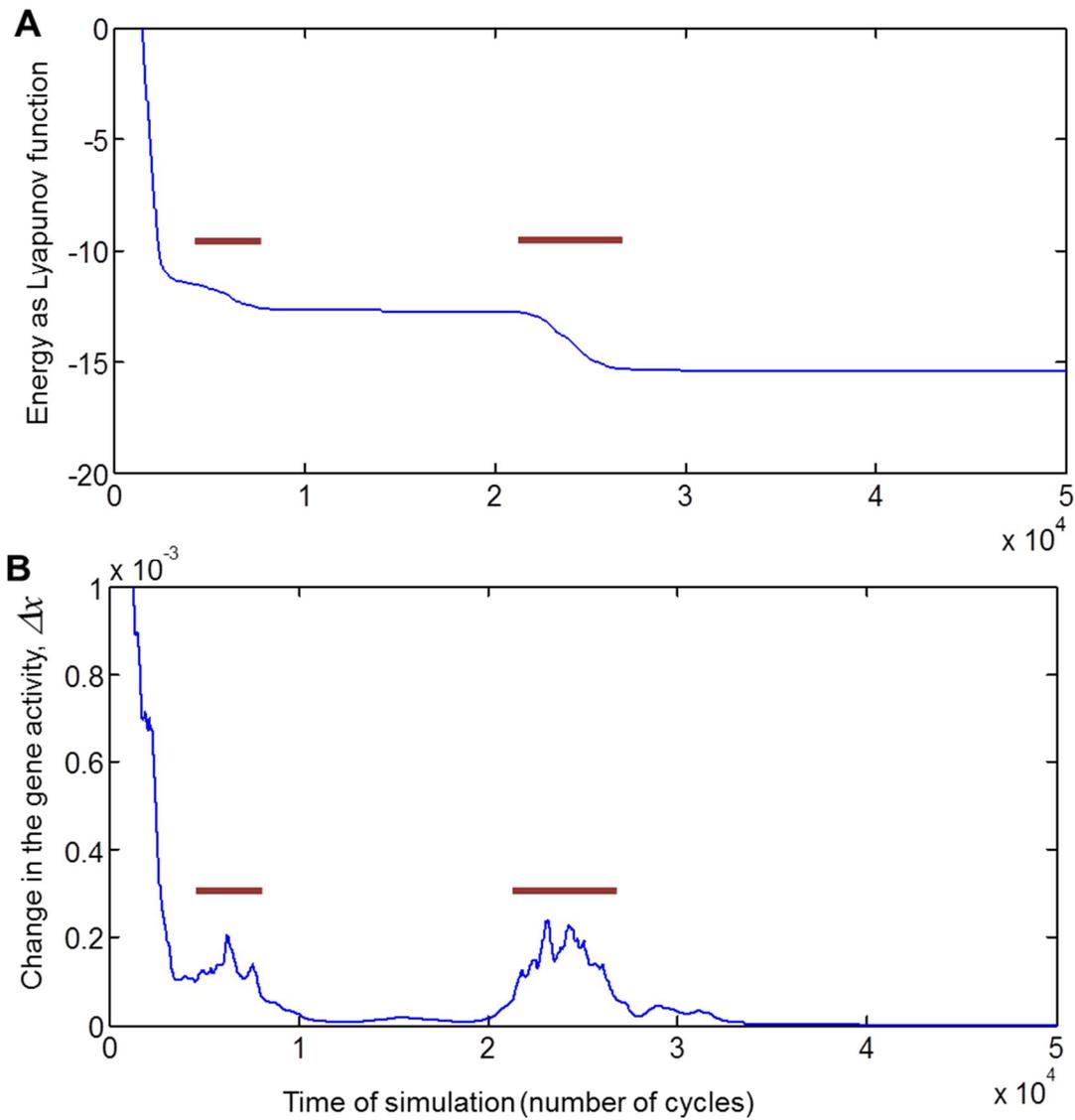

**Supplementary Figure 3.** The convergence to the spurious state is not gradual and is punctuated by sudden changes in the network state separated by the long periods of relative stability. The results are presented for the convergence to the mixture of 50% Plane and 50% Key. (A) The Lyapunov function always decreases over time. The decrease is not uniform and consists in slow decay punctuated by fast drops (brown bar). (B) The change in the gene activity vector, $\Delta x$, defined in supplementary methods, displays spikes that coincide with the drops in the Lyapunov function.